\documentclass[preprint]{aastex}

\begin{document}

\title {Long-term CCD Photometry and Physical Properties of the sdB+M Eclipsing System 2M 1533+3759 }
\author{Jae Woo Lee$^{1,2}$, Jae-Hyuck Youn$^1$, Kyeongsoo Hong$^1$, and Wonyong Han$^{1,2}$}
\affil{$^1$Korea Astronomy and Space Science Institute, Daejeon 34055, Korea}
\affil{$^2$Astronomy and Space Science Major, Korea University of Science and Technology, Daejeon 34113, Korea}
\email{jwlee@kasi.re.kr, jhyoon@kasi.re.kr, kshong@kasi.re.kr, whan@kasi.re.kr}

\begin{abstract}
New CCD photometry of seven successive years from 2010 is presented for the HW Vir-type eclipsing binary 2M 1533+3759. Using 
the $VI$ light curves together with the radial-velocity data given by For et al. (2010), we determined the absolute parameters 
of each component to be $M_1$ = 0.442$\pm$0.012 M$_\odot$, $M_2$ = 0.124$\pm$0.005 M$_\odot$, $R_1$ = 0.172$\pm$0.002 R$_\odot$, 
$R_2$ = 0.157$\pm$0.002 R$_\odot$, $L_1$ = 19.4$\pm$1.4 L$_\odot$, and $L_2$ = 0.002$\pm$0.002 L$_\odot$. These indicate that 
2M 1533+3759 is a detached system consisting of a normal sdB primary and an M7 dwarf companion. Detailed analyses of 
377 minimum epochs, including our 111 timings, showed that the orbital period of the system remains constant during the past 
12 yrs. Inspecting both types of minima, we found a delay of 3.9$\pm$1.0 s in the arrival times of the secondary eclipses 
relative to the primary eclipse times. This delay is in satisfactory agreement with the predicted R{\o}mer delay of 2.7$\pm$1.4 s 
and the result is the second measurement in sdB+M eclipsing binaries. The time shift of the secondary eclipse can be explained 
by some combination of the R{\o}mer delay and a non-zero eccentricity. Then, the binary star would have a very small eccentricity 
of $e \cos {\omega} \simeq$ 0.0001. 
\end{abstract}

{}

\section{INTRODUCTION}

Subdwarf B (sdB) stars are a class of hot ($T_{\rm eff}$ = 22,000$-$40,000 K) and compact ($\log$ $g$ = 5.0$-$6.2) stars 
with helium burning cores and hydrogen envelopes too thin to sustain hydrogen shell burning (Heber 2016). 
In the Hertzsprung-Russell diagram, they are located between the upper main sequence and the white dwarf sequence, also 
known as the so-called extreme horizontal branch. Recent studies have indicated that about half of the sdB stars are in 
close binaries with short periods of $P<$ 10 d (Maxted et al. 2001; Copperwheat et al. 2011), which are thought to be 
the products of common envelope evolution (Paczynski 1976; Han et al, 2003). The HW Vir-type stars, showing 
a striking reflection effect in light curves, are binary systems with orbital periods of 2$-$5 h and very low-mass companions 
(main sequence or brown dwarf). These binaries play an important role in establishing the mass distribution of the sdB stars, 
which is the key to understand their evolution. The mean mass of sdB stars derived from the binary modeling is 0.471 M$_\odot$, 
and it is in good agreement with 0.470 M$_\odot$ obtained from the asteroseismology of pulsating stars (Fontaine et al. 2012). 

Because the HW Vir systems are very short-period binaries with sharp eclipses and there are no peculiar light variations 
with time, we can measure the mid-eclipse times with timing accuracies of a few seconds from their light curves. 
The timing measurements are useful for understanding various astrophysical phenomena of binary stars (cf. Kreiner et al. 2001).
The presence of a circumbinary object orbiting a binary can cause a sinusoidal variation due to the light-travel-time (LTT) 
effect (Irwin 1952, 1959) in eclipse timing $O$--$C$ residuals, which are the differences between the observed ($O$) and 
the calculated ($C$) minima. If the timings with accuracies better than $\sim$10 s are enough to investigate 
the binary's period behavior, we can detect substellar-mass circumbinary companions such as planets and brown dwarfs 
(Ribas 2006; Lee et al. 2009a). Zorotovic \& Schreiber (2013) indicated that about 38 \% of the HW Vir binaries display 
orbital period changes that could be produced by substellar companions. Such systems are of great interest because they offer 
significant information about the formation and evolution of the post common-envelope binaries and their circumbinary objects. 

In order to look for substellar companions around the HW Vir binaries and to understand their physical properties, we choose 
2M 1533+3759 (NSVS 07826147; 2MASS J15334944+3759282; $V$ = $+$12.96). The program target was discovered by Kelley \& Shaw (2007) 
to be a potential sdB binary with an orbital period of about 3.88 h and a narrow eclipse width. For et al. (2010) obtained 
the $BVRI$ light curves and single-lined radial-velocity (RV) data. From separate analyses of the two datasets, they reported 
that the eclipsing binary is a detached system with a mass ratio of $q$ = 0.301, an inclination of $i$ = 86.6 deg, 
a surface gravity of $\log g_1$ = 5.58, and effective temperatures of $T_1$ = 29,230 K and $T_2$ = 3100 K. Individual masses 
and radii of the sdB primary and cool secondary were determined to be $M_1$ = 0.376 M$_\odot$, $M_2$ = 0.113 M$_\odot$, 
$R_1$ = 0.166 R$_\odot$, and $R_2$ = 0.152 R$_\odot$. The vast majority of sdB stars are clustered near the canonical mass 
of 0.47 M$_\odot$ (Fontaine et al. 2012), whereas the sdB mass in the binary system is unusually low. This paper is the fourth 
of a series of studies on the HW Vir-type systems (Lee et al. 2009a, 2014; Hong et al. 2017). We present the physical properties 
of 2M 1533+3759 from both the light-curve synthesis and the eclipse timing analysis, which are based on our new long-term observations.

\section{NEW LONG-TERM CCD PHOTOMETRY}

We took CCD photometric observations of 2M 1533+3759 for 76 nights from 2010 May through 2016 May, using the 1.0-m reflector 
at Mt. Lemmon Optical Astronomy Observatory (LOAO) in Arizona, USA. The observations of 2010 were made with a FLI IMG4301E 
2K CCD camera, which has 2084$\times$2084 pixels and a field of view (FOV) of 22.2$\times$22.2 arcmin$^2$. The observations 
of the other seasons were made with an ARC 4K CCD camera, which has 4096$\times$4096 pixels and a FOV of 28$\times$28 arcmin$^2$. 
We set up the 2$\times$2 binning modes for both cameras and their readout times are about 14 s including pre-flushing. 
A summary of the observations is given in Table 1, where we present the observing intervals, number of nights, filters, 
typical exposure times, number of observed points, and number of primary and secondary eclipses. The instruments and 
reduction methods for the CCD cameras are the same as those described by Lee et al. (2009b, 2012). 

In order to construct an artificial comparison star that would be optimal for the LOAO data, we monitored tens of stars imaged 
on the chip at the same time as 2M 1533+3759. Among them, four useful nearby stars were selected and combined by a weighted 
average. From all observing seasons, a total of 12,696 individual observations were obtained in the two bandpasses 
(3408 in $V$ and 9288 in $I$), and a sample of them is listed in Table 2, where the times are Barycentric Julian Dates (BJD) 
in the Barycentric Dynamical Time system (Eastman et al. 2010). The differential magnitudes from the artificial reference star 
were computed and the resultant light curves are displayed in Figure 1. Each light curve is labeled with the observing 
season and filter used (e.g., 2010$V$).

\section{LIGHT AND VELOCITY SOLUTIONS}

As shown in Figure 1, the light curves of 2M 1533+3759 display sharp eclipses and strong reflection effects, which increase 
towards longer wavelengths. Nonetheless, there are no significant differences among the seven seasonal datasets. The depths 
of the primary and secondary eclipses are 1.35 mag and 0.12 mag in the $V$ band, and 1.34 mag and 0.17 mag in the $I$ band. 
In order to obtain a consistent set of the binary parameters, we simultaneously modeled our long-term photometric data and 
the single-lined RV curve of For et al. (2010) by using the 2007 version of the Wilson-Devinney synthesis code 
(Wilson \& Devinney 1971; van Hamme \& Wilson 2007; hereafter W-D). The light-curve synthesis was performed in a way similar 
to that of the sdB+M eclipsing system HW Vir (Lee et al. 2009a). The method of multiple subsets (Wilson \& Biermann 1976) 
was used to determine the parameters and investigate our solution's stability. 

In our modeling, the surface temperature of the hot primary star was given as $T_{1}$ = 29,230 K from the spectroscopic 
analysis of For et al. (2010). The gravity-darkening exponents were assumed to be standard values of $g_1$=1.0 and $g_2$=0.32 
(von Zeipel 1924; Lucy 1967), while the bolometric albedos were fixed at $A_1$=$A_2$=1.0 (Rucinski 1969a,b) because the cool 
secondary star is highly irradiated and heated by the sdB primary. The logarithmic bolometric ($X$, $Y$) and monochromatic 
($x$, $y$) limb-darkening coefficients were initialized from the values of van Hamme (1993) in concert with the model atmosphere 
option. Furthermore, a synchronous rotation for both components was adopted and the detailed reflection effect was used 
(Wilson 1993). This synthesis was repeated until the correction of each free parameter became smaller than its standard error 
using the differential correction program of the W-D code. In this paper, the subscripts 1 and 2 refer to the sdB star and 
its companion, respectively. 

The mass ratio ($q$=$M_2$/$M_1$) is one of the most important parameters needed to understand the physical properties of binary 
systems. However, since there is still no spectroscopic mass ratio for 2M 1533+3759, we conducted a $q$-search procedure for 
a series of models with varying $q$ between 0.22 and 0.35, which correspond to the sdB mass range from $\sim$0.30 M$_\odot$ 
to $\sim$0.80 M$_\odot$ theoretically predicted by Han et al. (2003). The $q$ search was made simultaneously for all light and 
RV curves. In Figure 2, the weighted sum of the squared residuals ($\sum W(O-C)^2$; hereafter $\sum$) reached a global minimum 
at $q$=0.28, which was adopted as the initial value and thereafter adjusted to derive the binary parameters of the system. 
The final results are listed in the second and third columns of Table 3. We assumed that the temperatures of 
the primary and secondary components have the errors of 500 K and 600 K, respectively, following For et al. (2010). 
The synthetic $VI$ light curves are displayed as solid lines in Figure 3 and the synthetic RV curves are plotted in Figure 4. 
In the figures, our binary model appears to fit the light and RV curves quite well. At this point, we considered an orbital 
eccentricity of the binary star as an adjustable parameter but found that the value remained zero. The result implies that 
the eclipsing binary has negligible eccentricity. 

The consistent light and RV solution allows us to compute the absolute parameters of 2M 1533+3759 listed in Table 4, together 
with those of For et al. (2010) for comparison. The luminosity ($L$) and bolometric magnitudes ($M_{\rm bol}$) were obtained by 
adopting the solar values of $T_{\rm eff}$$_\odot$ = 5,780 K and $M_{\rm bol}$$_\odot$ = +4.73. For the absolute visual magnitudes 
($M_{\rm V}$), we used the bolometric corrections (BCs) appropriate for the effective temperature of each component using 
the correlation between $\log T_{\rm eff}$ and BC (Torres 2010). Most parameters are in accord with those of For et al. (2010) 
within the limits of their errors. The mass and surface gravity of the primary component presented in this paper are well matched 
with the canonical sdB value of 0.47 $\pm$ 0.03 M$_\odot$ (Fontaine et al. 2012) and the spectroscopic parameter of 
$\log$ $g$ = 5.58 $\pm$ 0.03 (For et al. 2010), respectively. The physical parameters of the secondary star correspond to 
a spectral type of approximately M7V. These indicate that 2M 1533+3759 is a detached binary with the sdB primary star slightly 
larger than the M-type dwarf companion. 

With an apparent visual magnitude of $V$ = +12.96 $\pm$ 0.17 (Kupfer et al. 2015) and the interstellar absorption of 
$A_{\rm V}$ = 0.035 (Schlafly \& Finkbeiner 2011), we derived the distance of the system to be 524 $\pm$ 47 pc. Compared 
with the result of For et al. (2010), we obtained a slightly higher luminosity but placed the binary star at closer distance. 
This may be mainly caused by the values of $V$ and $A_{\rm V}$ different from each other. For et al. (2010) appears to use 
$V$ = +13.61 (Kelly \& Shaw 2007) for a distance determination. Using their $V$ magnitude and our values, the distance was 
calculated to be 707 $\pm$ 63 pc, which is consistent with that (644 $\pm $66 pc) of For et al. (2010) within the errors. 
On the other hand, considering the temperature error of the primary star, we performed the light-curve synthesis for 28,730 K 
and 29,730 K. The light and RV parameters from both values are in good agreement with those from $T_1$ = 29,230 K. We can see 
that the adopted $T_1$ does not affect the physical parameters presented in this paper. 

Because the sdB primary component of 2M 1533+3759 lies at the boundary between both classes (V361 Hya and V1093 Her) of 
pulsating stars in the $T_{\rm eff}-\log g$ diagram (Green et al. 2011), it could be a candidate for hybrid pulsators. Using 
the PERIOD04 program (Lenz \& Breger 2005), we applied multiple frequency analyses to the light residuals from our binary models, 
but detected no pulsating periodicity with signal-to-noise amplitude ratios larger than 4.0 (Breger et al. 1993).

\section{ECLIPSE TIMING VARIATION AND ITS IMPLICATIONS}

From all our CCD observations, we determined 111 eclipse times and their errors using the method of Kwee \& van Woerden 
(1956). These are listed in Table 5, wherein Min I and Min II represent the primary and secondary minima, respectively. 
In addition to our measurements, 266 CCD timings were collected from Drake et al. (2010), For et al. (2010), Zhu \& Qian (2010), 
Backhaus et al. (2012), and Lohr et al. (2014). The HJD times based on UTC were transformed into TDB-based BJD ones using 
the online applets\footnote{http://astroutils.astronomy.ohio-state.edu/time/} developed by Eastman et al. (2010). 
For ephemeris computations, weights were calculated as the inverse squares of the timing errors.

The orbital period of 2M 1533+3759 has already been studied several times. Backhaus et al. (2012) and Lohr et al. (2014) 
reported that there is no clear evidence for the period change, while Zhu et al. (2015, 2016) suggested that a small amplitude 
cyclic variation may exist with a period of 7.6 yr, implying the presence of a Jupiter mass planet in the system. First of all, 
we introduced all eclipse times into a linear least-squares fit and thus found an improved ephemeris, as follows:
\begin{equation}
C = \mbox{BJD}~2,456,021.8529058(20) + 0.16177045211(32).
\end{equation}
The parenthesized numbers are the 1$\sigma$-error values for the last digit of each ephemeris term. The eclipse timing $O$--$C$ 
diagram constructed with equation (1) is plotted in the upper panel of Figure 5, where the blue and red circles are the primary 
and secondary eclipses, respectively. To see if the eclipse timings represent a real and periodic variation, 
we applied a periodogram analysis to the timing residuals, but no credible periodicity was found. Moreover, because recent 
timings appear to display a period decrease, we fitted the minimum epochs to a parabolic ephemeris. The result indicated that 
the quadratic term is not significant. Thus, the orbital period of the binary system can be considered constant.

In Figure 5, the primary and secondary timing residuals seem not to agree with each other. In order to examine this 
discrepancy in detail, we computed the secondary eclipse times related to one half period after the primary eclipse times 
and then plotted the difference ($\Delta t_{\rm SE}$) between the observed and computed times of the secondary eclipses in 
the lower panel of Figure 5. As can be seen from the figure, the secondary times shifted from zero by the mean value of 
$\Delta t_{\rm SE}$ = 3.9$\pm$1.0 s. The time difference between both eclipses could be caused by the R{\o}mer delay 
($\Delta t_{\rm Rd}$) in the binary star with a mass ratio far from unity (Kaplan 2010; Lee et al. 2017): 
\begin{equation}
 \Delta t_{\rm Rd} = {{P K_1} \over {\pi c}} ({1 \over q} - 1),
\end{equation}
where $P$ is the binary period, $K_1$ is the sdB velocity semi-amplitude, c is the speed of light, and $q$ is the mass ratio. 
Using the parameters in Table 3, we got a time delay of $\Delta t_{\rm Rd}$ = 2.7$\pm$1.4 s, which is in good agreement with 
the observed delay of $\Delta t_{\rm SE}$ in the arrival times of the secondary eclipses relative to the the primary eclipses. 

Although 2M 1533+3759 has negligible eccentricity ($e$), it does not mean that the eclipsing system is in a circular orbit. 
Even in a case that the binary's orbit has a very small eccentricity, the observed delay of $\Delta t_{\rm SE}$ could be affected 
by the time shift of $\Delta t_{\rm e}$ in the secondary eclipse due to non-zero eccentricity: 
\begin{eqnarray}
\Delta t_{\rm e} \simeq {{2Pe} \over {\pi}} \cos {\omega} = \Delta t_{\rm SE} - \Delta t_{\rm Rd},
\end{eqnarray}
where $\omega$ is the argument of periastron. Using the equation (3), the orbital eccentricity of 2M 1533+3759 is calculated 
to be $e \cos {\omega} \simeq$ 0.0001. If $\Delta t_{\rm SE}$ is fully produced by an eccentricity, $e \cos {\omega} \simeq$ 
0.0004. 

Finally, we added the time shft of $\Delta t_{\rm SE}$ = 3.9 s to the secondary eclipses and then applied a linear 
least-squares fit to all minimum times as before. As a result, the reference epoch and period of the new ephemeris are 
calculated to be BJD 2,456,021.8529060 and 0.16177045212 d, respectively, which are essentially identical to those given in 
equation (1). This is because the root-mean-square (rms) scatter ($\sim$10 s) of the timing measurements is about 2.5 times 
larger than the observed time delay in the secondary eclipses.

\section{DISCUSSION AND CONCLUSIONS}

In this paper, we presented new CCD photometry of 2M 1533+3759 made for seven years from 2010 to 2016. The $VI$ light curves, 
which display sharp eclipses and prominent reflection effects, were solved simultaneously with the single-lined RV curve 
of For et al. (2010). Our light and velocity solutions represent that the eclipsing pair is a HW Vir-type detached binary with 
a mass ratio of $q$ = 0.280, an orbital inclination of $i$ = 86.80 deg, and a temperature ratio between the components of 
$T_{2}/T_{1}$ = 0.106. The primary and secondary components fill $f_1$ = 39 \% and $f_2$ = 76 \% of their limiting lobe, 
respectively, where the filling factor $f_{1,2} = \Omega_{\rm in} / \Omega_{1,2}$. The masses and radii of both components are 
determined to be $M_1$ = 0.442 M$_\odot$, $M_2$ = 0.124 M$_\odot$, $R_1$ = 0.172 R$_\odot$, and $R_2$ = 0.157 R$_\odot$. 
These indicate that the primary is typical for a normal sdB star and the secondary is an M7 main sequence star. 

The eclipse times of 2M 1533+3759, spanning about 12 yrs, have been examined and they indicate that the orbital period 
has been essentially constant. From the time difference between both types of minima, we measured a delay of 3.9$\pm$1.0 s in 
the arrival times of the secondary eclipses relative to the primary eclipses. The observed value is in satisfactory accord with 
the expected R{\o}mer delay of 2.7$\pm$1.4 s across the binary orbit. This indicates that the R{\o}mer delay is the main cause 
of the time shift in the secondary eclipses measured in this paper, which is the second detection in HW Vir-type binaries 
(Barlow et al. 2012). Nonetheless, the time delay of the secondary eclipse may come from both the R{\o}mer delay and the non-zero 
eccentricity. Our result limits the eccentricity of 2M 1533+3759 to $e \le 0.0001$. Because the predicted R{\o}mer delay is less 
than $\sim$3 s, future timing measurements with accuracies better than the value will help to reveal more detailed properties of 
the binary star.

\acknowledgments{ }
The authors wish to thank the staff of LOAO for assistance during our observations. We appreciate the careful reading and 
valuable comments of the anonymous referee. This research has made use of the Simbad database maintained at CDS, Strasbourg, 
France, and was supported by the KASI grant 2017-1-830-03. The work by K. Hong and W. Han was supported by grant numbers 
NRF-2016R1A6A3A01007139 and NRF-2014M1A3A3A02034746 of the National Research Foundation of Korea (NRF), respectively.

\clearpage
\begin{figure}
\includegraphics[scale=0.85]{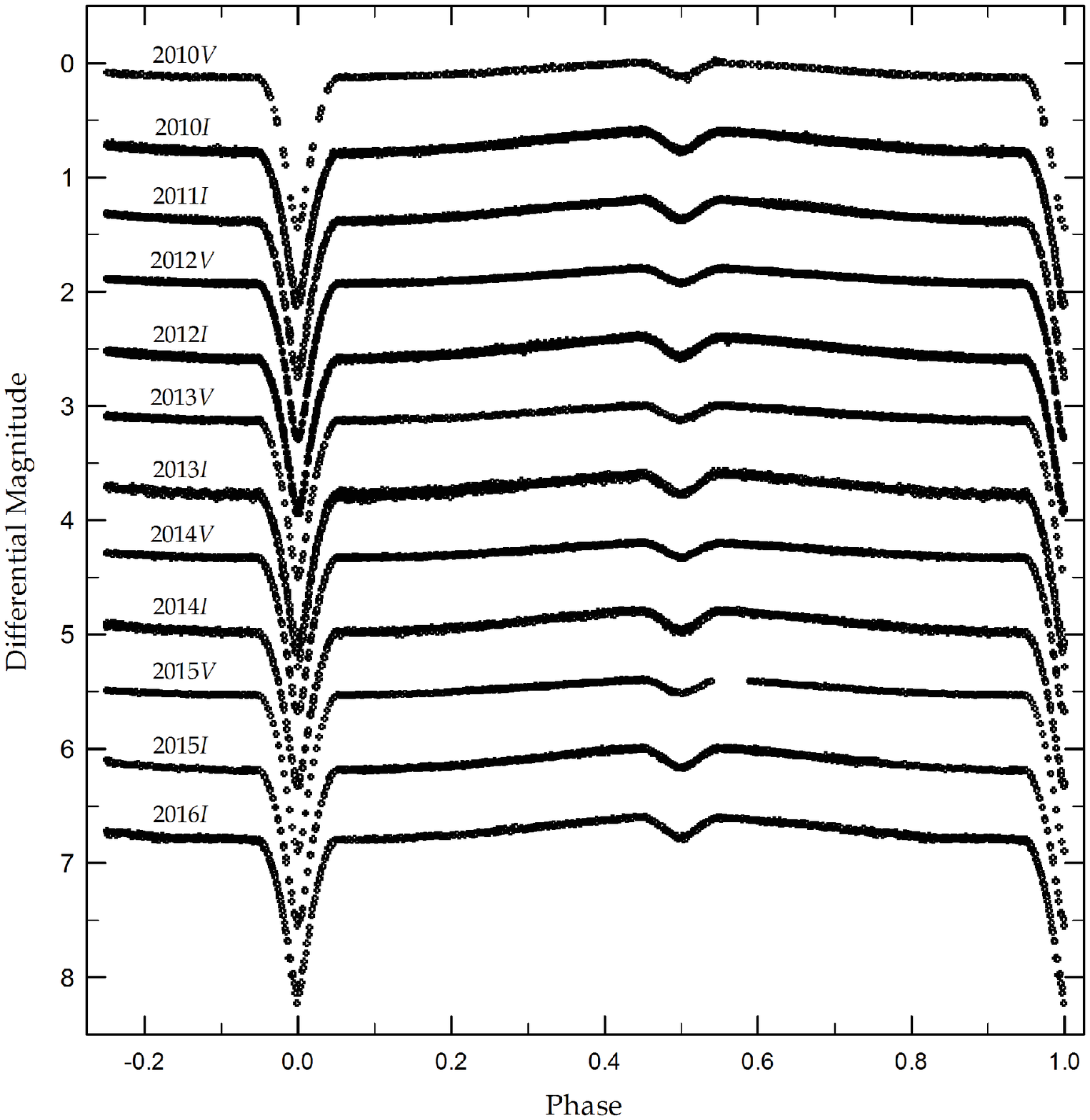}
\caption{Light curves of 2M 1533+3759 obtained from 2010 to 2016 in $VI$ bandpasses. All but 2010$V$ are displaced vertically for clarity. }
\label{Fig1}
\end{figure}

\begin{figure}
\includegraphics[]{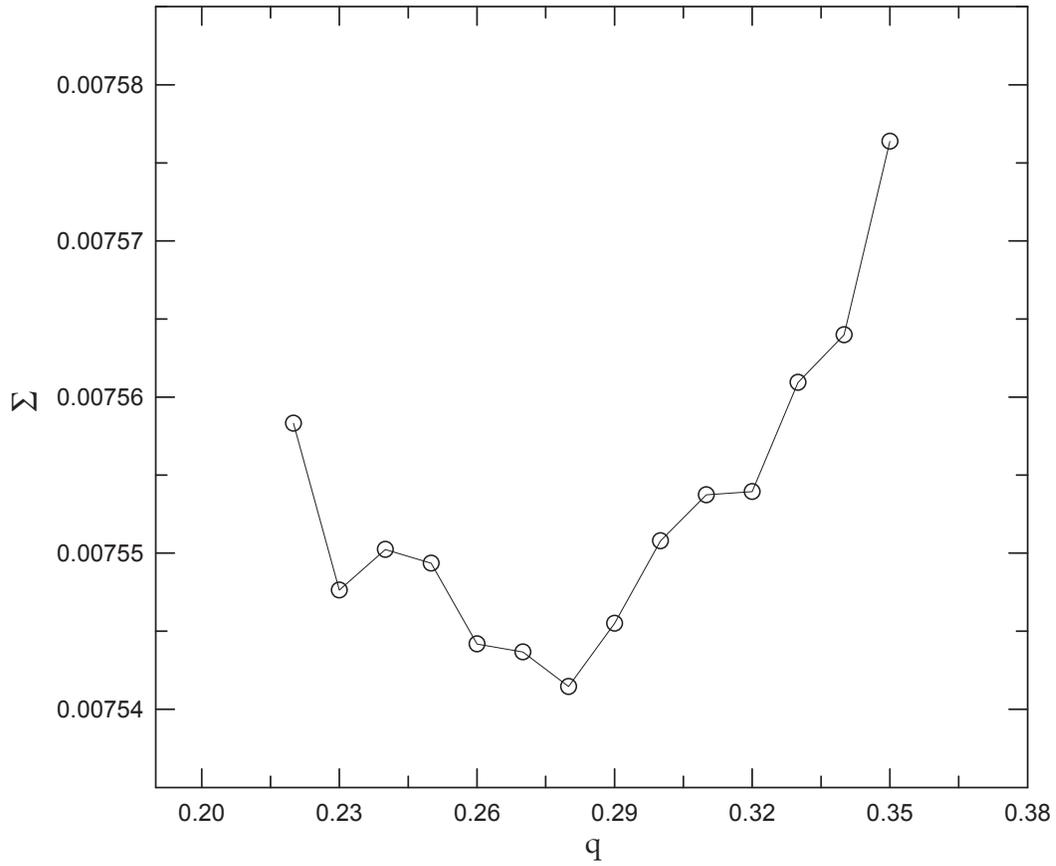}
\caption{Behavior of $\sum$ of 2M 1533+3759 as a function of mass ratio $q$, showing a global minimum at $q$=0.28. }
\label{Fig2}
\end{figure}

\begin{figure}
\includegraphics[]{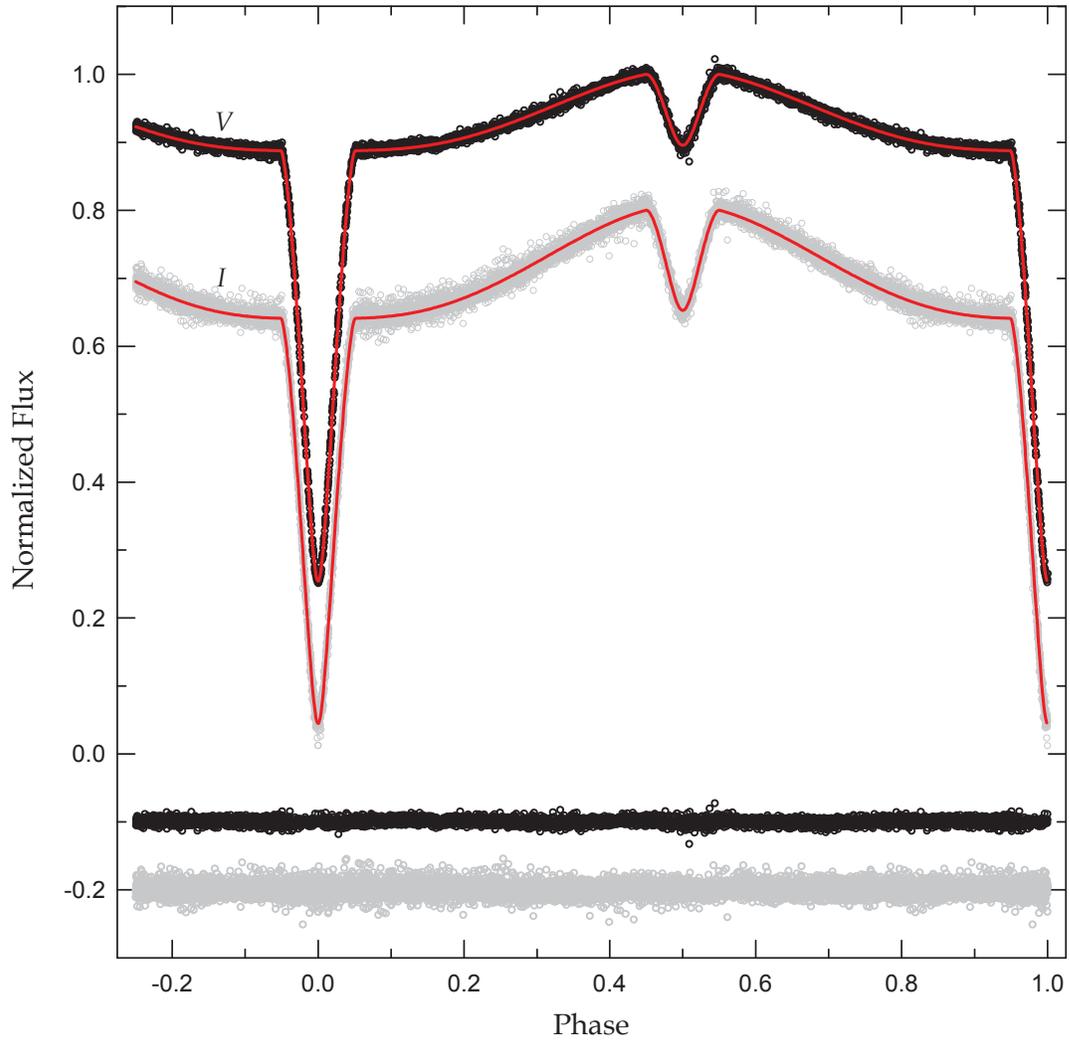}
\caption{Normalized $V$ (black circle) and $I$ (gray circle) light curves with fitted models. The red solid curves are computed 
with the model parameters of Table 3. The corresponding residuals from the fits are offset from zero and plotted at the bottom in 
the same order as the light curves. }
\label{Fig3}
\end{figure}

\begin{figure}
\includegraphics[]{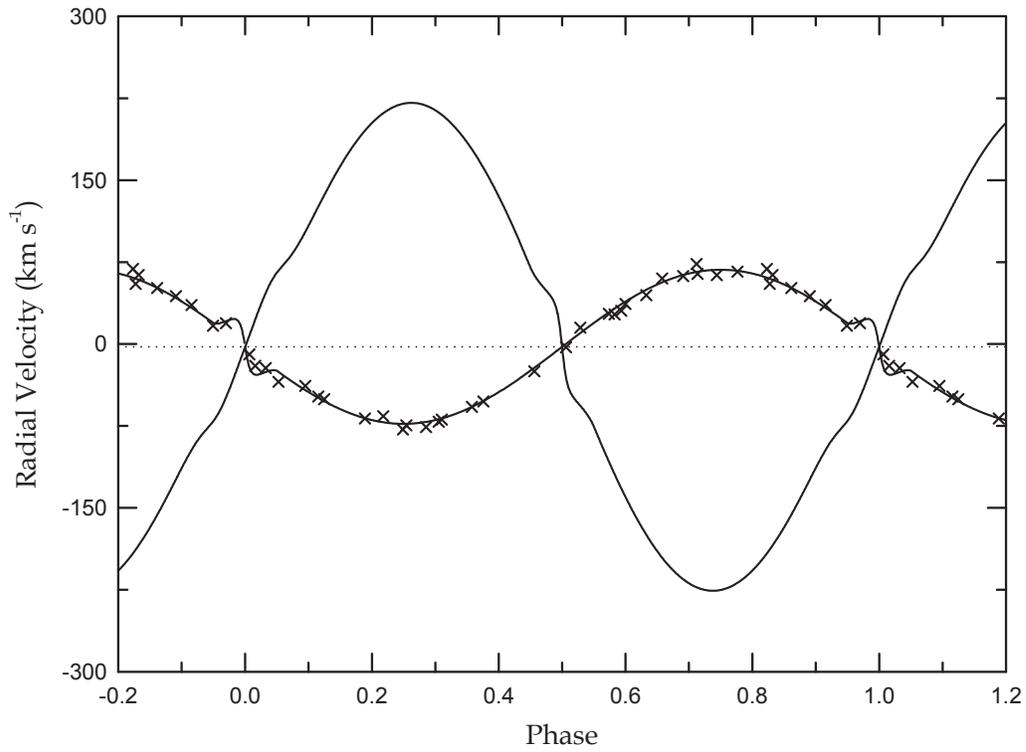}
\caption{Radial-velocity curves of 2M 1533+3759. The `x' symbols are the measurements of For et al. (2010), while the solid curves 
denote the result from consistent light and RV curve analysis. The dotted line refers to the systemic velocity of $-$2.5 km s$^{-1}$. }
\label{Fig4}
\end{figure}

\begin{figure}
\includegraphics[]{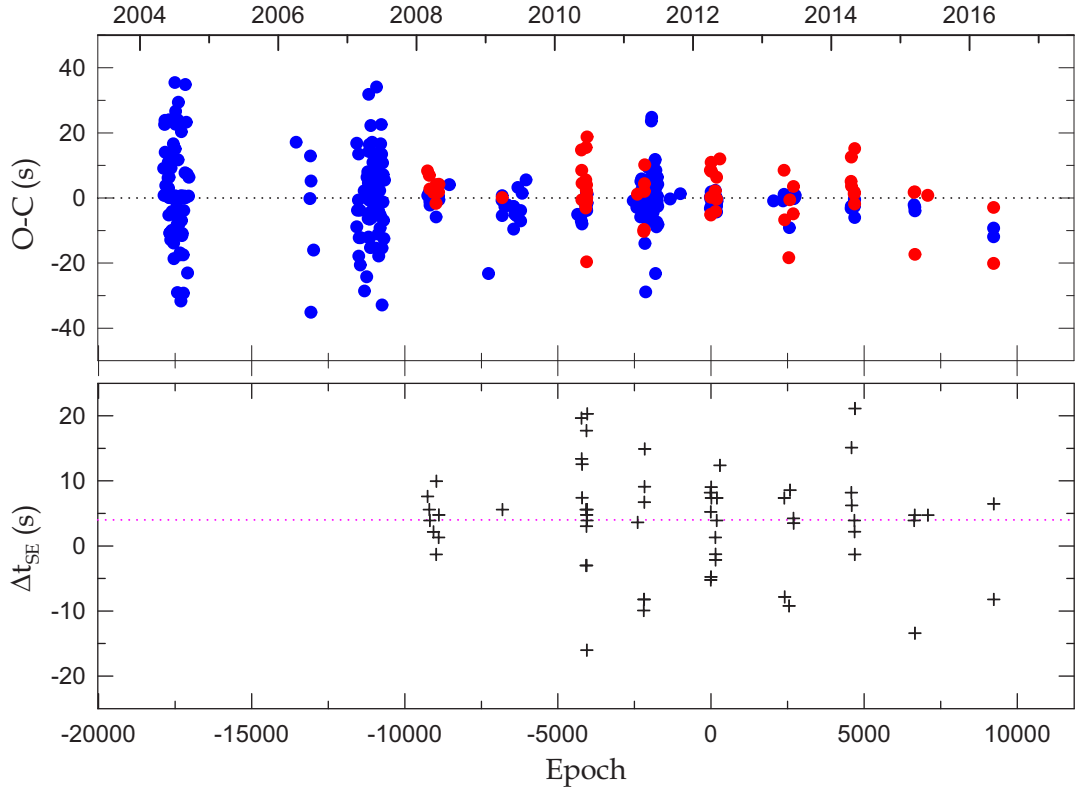}
\caption{The upper panel shows the eclipse timing diagram of 2M 1533+3759 constructed with the linear ephemeris of equation (1).  
The blue and red circles represent the primary and secondary minima, respectively. The lower panel displays the time delay 
($\Delta t_{\rm SE}$) of the secondary eclipses related to one half period after the primary eclipse times. The dotted line 
refers to the mean value of $\Delta t_{\rm SE}$ = 3.9 s. }
\label{Fig5}
\end{figure}

\clearpage
\begin{deluxetable}{llccccc}
\tablewidth{0pt}
\tablecaption{Observing log of 2M 1533+3759.}
\tablehead{
\colhead{Season} & \colhead{Observing Interval} & \colhead{$N_{\rm night}$} & \colhead{Filter} & \colhead{Exp. Time (s)} & \colhead{$N_{\rm obs}$} & \colhead{$N_{\rm MinI/MinII}$}
}                                                                                                                                                                            
\startdata                                                                                                                                                                   
2010             & May 01$-$June 19             & 12                        & $VI$             & 70                      &  2502                   & 16/14                         \\  
2011             & March 10$-$April 20          & 12                        & $I$              & 55                      &  1463                   & 7/7                           \\  
2012             & March 30$-$May 22            & 20                        & $VI$             & 70                      &  3104                   & 16/12                         \\  
2013             & March 01$-$June 24           & 19                        & $VI$             & 45                      &  2073                   & 9/6                           \\  
2014             & April 15$-$May 04            & 7                         & $VI$             & 35                      &  1929                   & 6/7                           \\  
2015             & March 13$-$May 24            & 5                         & $VI$             & 35                      &  1016                   & 3/4                           \\  
2016             & May 06                       & 1                         & $I$              & 35                      &  609                    & 2/2                           \\ [1.0mm]
\enddata
\end{deluxetable}

\begin{deluxetable}{lclc}
\tablewidth{0pt} 
\tablecaption{CCD photometric observations of 2M 1533+3759.}
\tablehead{
\colhead{BJD}    & \colhead{$\Delta V$ (mag)} & \colhead{BJD}    & \colhead{$\Delta I$ (mag)} 
}
\startdata                                                       
2,455,360.65054  & 0.0684                     & 2,455,317.93713  & 0.0675                     \\
2,455,360.65244  & 0.0712                     & 2,455,317.93929  & 0.0699                     \\
2,455,360.65432  & 0.0792                     & 2,455,317.94036  & 0.0731                     \\
2,455,360.65621  & 0.0851                     & 2,455,317.94144  & 0.0834                     \\
2,455,360.65810  & 0.0945                     & 2,455,317.94246  & 0.0863                     \\
2,455,360.65998  & 0.0986                     & 2,455,317.94342  & 0.0942                     \\
2,455,360.66187  & 0.1000                     & 2,455,317.94438  & 0.0944                     \\
2,455,360.66377  & 0.1008                     & 2,455,317.94534  & 0.1032                     \\
2,455,360.66565  & 0.1114                     & 2,455,317.94630  & 0.0956                     \\
2,455,360.66754  & 0.1101                     & 2,455,317.94726  & 0.1118                     \\ [1.0mm]
\enddata
\tablecomments{This table is available in its entirety in machine-readable form. }
\end{deluxetable}

\begin{deluxetable}{lcc}
\tablewidth{0pt}
\tablecaption{Velocity and light curve parameters of 2M 1533+3759.}
\tablehead{
\colhead{Parameter}                      & \colhead{Primary} & \colhead{Secondary}
}
\startdata                                                                         
$T_0$ (HJD)                              & \multicolumn{2}{c}{2,456,021.8528710$\pm$0.0000076}        \\
$P$ (day)                                & \multicolumn{2}{c}{0.1617704495$\pm$0.0000000021}          \\
$a$ (R$_\odot$)                          & \multicolumn{2}{c}{1.033$\pm$0.017}                        \\
$\gamma$ (km s$^{-1}$)                   & \multicolumn{2}{c}{$-$2.5$\pm$1.1}                         \\
$K_1$ (km s$^{-1}$)                      & \multicolumn{2}{c}{70.6$\pm$2.1}                           \\
$K_2$ (km s$^{-1}$)                      & \multicolumn{2}{c}{252.0$\pm$3.2}                          \\
$q$                                      & \multicolumn{2}{c}{0.2802$\pm$0.0049}                      \\
$i$ (deg)                                & \multicolumn{2}{c}{86.802$\pm$0.066}                       \\
$T$ (K)                                  & 29230$\pm$500             & 3089$\pm$600                   \\
$\Omega$                                 & 6.289$\pm$0.044           & 3.208$\pm$0.010                \\
$\Omega_{\rm in}$                        & \multicolumn{2}{c}{2.422}                                  \\
$X$, $Y$                                 & 0.762, 0.255              & 0.463, 0.290                   \\
$x_{V}$, $y_{V}$                         & 0.334$\pm$0.043, 0.210    & 0.855$\pm$0.084, 0.361         \\
$x_{I}$, $y_{I}$                         & 0.261$\pm$0.058, 0.164    & 0.748$\pm$0.073, 0.352         \\
$L/(L_1+L_2)_{V}$                        & 0.9998$\pm$0.0002         & 0.0002$\pm$0.0001              \\
$L/(L_1+L_2)_{I}$                        & 0.9976$\pm$0.0005         & 0.0024$\pm$0.0004              \\
$r$ (pole)                               & 0.1663$\pm$0.0012         & 0.1507$\pm$0.0006              \\
$r$ (point)                              & 0.1672$\pm$0.0012         & 0.1553$\pm$0.0007              \\
$r$ (side)                               & 0.1668$\pm$0.0012         & 0.1519$\pm$0.0006              \\
$r$ (back)                               & 0.1671$\pm$0.0012         & 0.1546$\pm$0.0006              \\
$r$ (volume)$\rm ^a$                     & 0.1668$\pm$0.0012         & 0.1525$\pm$0.0006              \\
\enddata
\tablenotetext{a}{Mean volume radius computed from the tables given by Mochnacki (1984). }
\end{deluxetable}

\begin{deluxetable}{lccccc}
\tablewidth{0pt} 
\tablecaption{Absolute parameters for 2M 1533+3759.}
\tablehead{
\colhead{Parameter}   & \multicolumn{2}{c}{For et al. (2010)}      && \multicolumn{2}{c}{This Work}               \\ [1.0mm] \cline{2-3} \cline{5-6} \\[-2.0ex]
                      & \colhead{Primary} & \colhead{Secondary}    && \colhead{Primary}  & \colhead{Secondary}            
}
\startdata 
$M$ (M$_\odot$)       & 0.376$\pm$0.055    & 0.113$\pm$0.017       && 0.442$\pm$0.012    & 0.124$\pm$0.005        \\          
$R$ (R$_\odot$)       & 0.166$\pm$0.007    & 0.152$\pm$0.005       && 0.172$\pm$0.002    & 0.157$\pm$0.002        \\          
$\log$ $g$ (cgs)      & 5.58$\pm$0.03      & \dots                 && 5.61$\pm$0.02      & 5.14$\pm$0.02          \\          
$L$ (L$_\odot$)       & 18.14$\pm$1.84     & \dots                 && 19.4$\pm$1.4       & 0.002$\pm$0.002        \\          
$M_{\rm bol}$ (mag)   & \dots              & \dots                 && $+$1.51$\pm$0.08   & $+$11.47$\pm$0.84      \\          
BC (mag)              & \dots              & \dots                 && $-$2.82$\pm$0.05   & $-$4.36$\pm$4.71       \\          
$M_{\rm V}$ (mag)     & $+$4.57$\pm$0.21   & \dots                 && $+$4.33$\pm$0.09   & $+$15.83$\pm$4.78      \\          
Distance (pc)         & \multicolumn{2}{c}{644$\pm$66}             && \multicolumn{2}{c}{524$\pm$47}              \\          
\enddata
\end{deluxetable}

\begin{deluxetable}{lccclccc}
\tablewidth{0pt}
\tablecaption{New times of minimum light for 2M 1533+3759. }
\tablehead{
BJD          &  Error         & Filter  &  Min    &   BJD          &   Error        & Filter  &  Min     \\
(2,450,000+) &                &         &         &   (2,450,000+) &                &         &          }
\startdata
5317.98961   &  $\pm$0.00002  &   $I$   &   I     &   6044.74345   &  $\pm$0.00003  &   $V$   &   II     \\
5336.91673   &  $\pm$0.00002  &   $I$   &   I     &   6044.82432   &  $\pm$0.00002  &   $V$   &   I      \\
5337.96849   &  $\pm$0.00003  &   $I$   &   II    &   6045.71405   &  $\pm$0.00005  &   $I$   &   II     \\
5338.85800   &  $\pm$0.00002  &   $I$   &   I     &   6045.79496   &  $\pm$0.00001  &   $I$   &   I      \\
5338.93904   &  $\pm$0.00004  &   $I$   &   II    &   6045.87577   &  $\pm$0.00005  &   $I$   &   II     \\
5340.71841   &  $\pm$0.00006  &   $I$   &   II    &   6045.95667   &  $\pm$0.00002  &   $I$   &   I      \\
5340.88024   &  $\pm$0.00004  &   $I$   &   II    &   6046.76555   &  $\pm$0.00001  &   $V$   &   I      \\
5340.96098   &  $\pm$0.00002  &   $I$   &   I     &   6046.92729   &  $\pm$0.00002  &   $V$   &   I      \\
5359.88820   &  $\pm$0.00002  &   $I$   &   I     &   6051.78041   &  $\pm$0.00001  &   $I$   &   I      \\
5360.69704   &  $\pm$0.00007  &   $V$   &   I     &   6051.86132   &  $\pm$0.00004  &   $I$   &   II     \\
5360.85882   &  $\pm$0.00007  &   $V$   &   I     &   6051.94216   &  $\pm$0.00003  &   $I$   &   I      \\
5361.74854   &  $\pm$0.00005  &   $I$   &   II    &   6052.75104   &  $\pm$0.00001  &   $V$   &   I      \\
5361.82946   &  $\pm$0.00002  &   $I$   &   I     &   6052.83202   &  $\pm$0.00009  &   $V$   &   II     \\
5361.91041   &  $\pm$0.00008  &   $I$   &   II    &   6052.91282   &  $\pm$0.00001  &   $V$   &   I      \\
5363.69000   &  $\pm$0.00007  &   $I$   &   II    &   6053.88345   &  $\pm$0.00001  &   $I$   &   I      \\
5363.77068   &  $\pm$0.00002  &   $I$   &   I     &   6068.84736   &  $\pm$0.00006  &   $I$   &   II     \\
5363.85162   &  $\pm$0.00005  &   $I$   &   II    &   6352.99701   &  $\pm$0.00002  &   $I$   &   I      \\
5363.93247   &  $\pm$0.00001  &   $I$   &   I     &   6402.98408   &  $\pm$0.00001  &   $V$   &   I      \\
5364.66049   &  $\pm$0.00005  &   $I$   &   II    &   6408.80783   &  $\pm$0.00002  &   $I$   &   I      \\
5364.74132   &  $\pm$0.00001  &   $I$   &   I     &   6408.88881   &  $\pm$0.00008  &   $I$   &   II     \\
5364.82220   &  $\pm$0.00004  &   $I$   &   II    &   6408.96961   &  $\pm$0.00003  &   $I$   &   I      \\
5364.90312   &  $\pm$0.00001  &   $I$   &   I     &   6410.99165   &  $\pm$0.00008  &   $V$   &   II     \\
5365.71191   &  $\pm$0.00003  &   $V$   &   I     &   6433.96292   &  $\pm$0.00005  &   $I$   &   II     \\
5365.79261   &  $\pm$0.00007  &   $V$   &   II    &   6436.95578   &  $\pm$0.00005  &   $I$   &   I      \\
5366.76346   &  $\pm$0.00007  &   $V$   &   II    &   6438.97801   &  $\pm$0.00008  &   $I$   &   II     \\
5366.84430   &  $\pm$0.00003  &   $V$   &   I     &   6456.93448   &  $\pm$0.00007  &   $I$   &   II     \\
5366.92525   &  $\pm$0.00007  &   $V$   &   II    &   6457.90520   &  $\pm$0.00007  &   $V$   &   II     \\
5367.65321   &  $\pm$0.00003  &   $I$   &   I     &   6459.92729   &  $\pm$0.00001  &   $V$   &   I      \\
5367.81495   &  $\pm$0.00002  &   $I$   &   I     &   6460.89791   &  $\pm$0.00003  &   $I$   &   I      \\
5367.89607   &  $\pm$0.00007  &   $I$   &   II    &   6465.91280   &  $\pm$0.00005  &   $I$   &   I      \\
5631.01546   &  $\pm$0.00001  &   $I$   &   I     &   6466.88343   &  $\pm$0.00002  &   $I$   &   I      \\
5632.95671   &  $\pm$0.00002  &   $I$   &   I     &   6762.92331   &  $\pm$0.00002  &   $I$   &   I      \\
5634.00826   &  $\pm$0.00006  &   $I$   &   II    &   6763.00429   &  $\pm$0.00007  &   $I$   &   II     \\
5662.88427   &  $\pm$0.00001  &   $I$   &   I     &   6763.89394   &  $\pm$0.00001  &   $V$   &   I      \\
5666.84753   &  $\pm$0.00005  &   $I$   &   II    &   6763.97500   &  $\pm$0.00010  &   $V$   &   II     \\
5666.92853   &  $\pm$0.00002  &   $I$   &   I     &   6764.94552   &  $\pm$0.00006  &   $I$   &   II     \\
5667.89914   &  $\pm$0.00002  &   $I$   &   I     &   6778.85775   &  $\pm$0.00003  &   $V$   &   II     \\
5667.97993   &  $\pm$0.00003  &   $I$   &   II    &   6778.93859   &  $\pm$0.00001  &   $V$   &   I      \\
5668.86976   &  $\pm$0.00001  &   $I$   &   I     &   6779.82838   &  $\pm$0.00004  &   $I$   &   II     \\
5668.95055   &  $\pm$0.00006  &   $I$   &   II    &   6779.90924   &  $\pm$0.00001  &   $I$   &   I      \\
5669.92131   &  $\pm$0.00003  &   $I$   &   II    &   6780.87986   &  $\pm$0.00001  &   $V$   &   I      \\
5670.89196   &  $\pm$0.00003  &   $I$   &   II    &   6780.96073   &  $\pm$0.00003  &   $V$   &   II     \\
5670.97274   &  $\pm$0.00001  &   $I$   &   I     &   6781.85042   &  $\pm$0.00001  &   $I$   &   I      \\
5671.86265   &  $\pm$0.00006  &   $I$   &   II    &   6781.93155   &  $\pm$0.00006  &   $I$   &   II     \\
6018.77927   &  $\pm$0.00003  &   $V$   &   I     &   7094.95722   &  $\pm$0.00005  &   $I$   &   II     \\
6018.86025   &  $\pm$0.00007  &   $V$   &   II    &   7095.03806   &  $\pm$0.00002  &   $I$   &   I      \\
6018.94100   &  $\pm$0.00003  &   $V$   &   I     &   7097.94992   &  $\pm$0.00001  &   $V$   &   I      \\
6020.80140   &  $\pm$0.00007  &   $I$   &   II    &   7098.03086   &  $\pm$0.00007  &   $V$   &   II     \\
6020.96311   &  $\pm$0.00003  &   $I$   &   II    &   7098.92053   &  $\pm$0.00001  &   $I$   &   I      \\
6021.85290   &  $\pm$0.00002  &   $I$   &   I     &   7099.00126   &  $\pm$0.00009  &   $I$   &   II     \\
6021.93373   &  $\pm$0.00004  &   $I$   &   II    &   7166.94506   &  $\pm$0.00004  &   $I$   &   II     \\
6022.82355   &  $\pm$0.00002  &   $I$   &   I     &   7514.67050   &  $\pm$0.00002  &   $I$   &   I      \\
6022.90454   &  $\pm$0.00006  &   $I$   &   II    &   7514.75129   &  $\pm$0.00007  &   $I$   &   II     \\
6022.98527   &  $\pm$0.00001  &   $I$   &   I     &   7514.83230   &  $\pm$0.00001  &   $I$   &   I      \\
6023.87513   &  $\pm$0.00003  &   $V$   &   II    &   7514.91326   &  $\pm$0.00006  &   $I$   &   II     \\
6023.95593   &  $\pm$0.00001  &   $V$   &   I     &                                                      \\
\enddata
\end{deluxetable}

\end{document}